\documentclass[showpacs,amsmath,amssymb, nobibnotes, aps, prl,showkeys, floatfix]{revtex4}
\usepackage{graphicx}
\usepackage{dcolumn}
\usepackage{bm}
\usepackage{docs}
\usepackage{amssymb}
\usepackage{amsmath}
\usepackage{epsfig}
\expandafter\ifx\csname package@font\endcsname\relax\else
 \expandafter\expandafter
 \expandafter\usepackage
 \expandafter\expandafter
 \expandafter{\csname package@font\endcsname}%
\fi

\newcommand{\ltsima} {$\; \buildrel < \over \sim \;$}
\newcommand{\gtsima} {$\; \buildrel > \over \sim \;$}
\newcommand{\lta} {\lower.5ex\hbox{\ltsima}}
\newcommand{\gta} {\lower.5ex\hbox{\gtsima}}

\newcommand{\RNum}[1]{\uppercase\expandafter{\romannumeral #1\relax}}

\begin{document}

\title[Newtonian like representation of spacetime geometries]
{Exact relativistic Newtonian representation of gravitational static spacetime geometries}

\author {Shubhrangshu Ghosh$^{1}$ \thanks{Email: sghosh@jcbose.ac.in}, Tamal Sarkar$^{2,3}$ \thanks{Email: ta.sa.nbu@hotmail.com}, Arunava Bhadra$^{2}$ \thanks{Email: 
aru\_bhadra@yahoo.com}}

\affiliation{ $^{1}$ Center for Astroparticle Physics and Space Science, Department of Physics, Bose Institute, Block EN, Sector V, Salt Lake, 
Kolkata, India 700091 }
\affiliation{ $^{2}$ High Energy $\&$ Cosmic Ray Research Centre, University of North Bengal, Post N.B.U, Siliguri, West Bengal, India 734013  }
\affiliation{ $^{3}$ University Science Instrumentation Center, University of North Bengal, Post N.B.U, Siliguri, West Bengal, India 734013  }
 



\begin{abstract}

We construct a self-consistent relativistic Newtonian analogue corresponding to gravitational static spherical symmetric spacetime geometries, staring directly from a generalized scalar relativistic gravitational action in Newtonian framework, which gives geodesic equations of motion identical to those of the parent metric. Consequently, the derived velocity-dependent relativistic scalar potential, which is a relativistic generalization of Newtonian gravitational potential, exactly reproduces the relativistic gravitational features corresponding to any static spherical symmetric spacetime geometry in its entirety, including all the experimentally tested gravitational effects in the weak field up to the present. This relativistic analogous potential is expected to be quite useful in studying wide range of astrophysical phenomena, especially in strong field gravity.




\end{abstract}

\maketitle	

\section{Introduction}

The Schiff conjecture [1] demands that any viable theory of gravity should respect Einstein equivalence principle which in turn implies that any feasible configuration of the gravitational field will correspond to a unique metric and the world lines of a test particle is the geodesics of that metric. Though no rigorous proof of the Schiff conjecture exists till now, the Eotvos-Dicke-Braginsky experiment and few other observations indicate the validity of the Schiff 
conjecture [2,3]
General relativity, which is widely considered as the standard theory of gravitation, is the metric theory of gravitation; the theory was constructed based on space time geometry without any notion of potential which was the cornerstone in Newtonian picture or Newtonian like description of the physical universe. Baring one or two, all the viable alternative theories of gravitation are also based on spacetime metric (geometry).

Newtonian kind of description (through potential) of gravity is, however, mathematically much more comprehensible, as well as simpler for understanding the concerned phenomena physically compared to geometrical (metric) description of gravity. As a result, Newtonian potential continues to be used in most applications at far (weak) field static condition where Newtonian approximation is reasonably valid. At strong gravity, however, Newtonian description is no longer valid, even approximately. Nonetheless, for simpler understanding of the general relativistic (GR) behavior of a test particle under the influence of a non-gravitation potential in a curved background spacetime, efforts have been made to formulate or prescribe modified Newtonian like potentials to mimic GR gravitational features of few corresponding spacetime geometries, particularly in the strong field gravity. Although not intended to construct Newtonian analogues of gravitational spacetime geometries, they are mainly endeavoured to mathematically replicate few GR features approximately, to be mostly used to study inner relativistic dynamics of the accretion flow around spacetime geometries describing black holes (BHs)/compact objects; commonly called pseudo-Newtonian potentials (PNPs). These potentials, although not exactly Newtonian in nature, are notionally Newtonian in character, in a sense that the gravitational interaction can be treated through the form of a scalar potential and a corresponding force law. Since its inception by Paczy\'nski \& Wiita (1980) [4] to mimic GR features of the spherically symmetric Schwarzschild geometry, many modified Newtonian like potentials have been proposed in the literature corresponding to different spacetime geometries. Nonetheless, most of these modified Newtonian like potentials which are either prescribed through an ad hoc proposition or formulated preserving certain specific procedures [5] lack the uniqueness to encompass most of the GR features simultaneously within a reasonable accuracy (for detailed analysis of various PNPs and their use see 
[6,7,8,9,10]). More importantly, they do not give the classical predictable GR effects or reproduce the classical experimental tests of general relativity like gravitational deflection of light, gravitational precession of test particles or gravitational time delay, even in the weak field regime which have been tested experimentally. Recently, a generic Paczy\'nski-Wiita like PNP corresponding to any static spherical BH (in Einstein's gravity as well as in modified gravity theories) has been 
formulated [11]. In general, the modifications of Newtonian gravitational potential are done in such a manner that they are emphasized to correctly reproduce the location of the inner most stable circular orbit (ISCO) and the marginally bound orbit of the test particle motion in the  strong field gravity, or at least approximately. 
Wegg (2012) [12] proposed couple of modified Newtonian like potentials to reproduce precessional effects in general relativity for orbits with large apoapsis, however, they are not quite effective in the vicinity of the Schwarzschild BH, in strong field gravity. 

In recent times, very accurate analogous potentials/modified Newtonian like potentials of corresponding GR geometries describing BHs/naked singularities have been derived from the conserved Hamiltonian of the test particle motion, comprising of a velocity dependent part of the particle motion along with the usual spherically symmetric part of the source and the other source dependent terms, and can be referred as some kind of Newtonian like analogous potentials (NAPs) [13,14,8,9,10]. Although many salient GR features can be reproduced with NAPs with precise/reasonable accuracy, however they fail to reproduce temporal effects like angular and epicyclic frequencies of test particle motion, accurately. Nonetheless, as compared to the existing PNPs/modified Newtonian like potentials, velocity dependent NAP of corresponding Schwarzschild geometry when used in simple astrophysical situations in strong field gravity,  
renders much better approximation to GR results [13]. Several other attempts have also been made to describe some form of Newtonian analogue of GR effects or to mimic GR effects with Newtonian dynamics [15,16].  

In the present work, we propose to construct an {\it exact relativistic Newtonian like representation} for the general static spherically symmetric spacetime geometry from a first principle approach. Such a construct should be essentially a relativistic Newtonian analogue described through a relativistic scalar potential and a corresponding force law, which then essentially demands that the corresponding modified Newtonian potential will give path equations of test particles exactly the same to those obtained from geodesic equations of the corresponding spacetime metric without any restriction on particle energy or the strength of gravitational field. Equivalently, it implies that if it can be demonstrated that any Newtonian analogous construct exactly reproduces the corresponding geodesic equations of motion of the corresponding spacetime geometry, then such a construct will essentially be an exact relativistic Newtonian analogous theory of gravitational spacetime geometries. The corresponding relativistic scalar potential will reproduce all the relativistic features of the corresponding spacetime geometry in toto including all the classical experimental tests of general relativity, while working in Newtonian framework. 

In the next section, we construct our proposed relativistic Newtonian analogous theory. Although our construct would be generic in nature, however, we would mainly focus on static spherically symmetric spacetime geometries. 

\section{Formulation of exact relativistic Newtonian analogous potential for static spacetime geometries}

Static spacetimes are among the simplest class of Lorentzian manifolds with a non-vanishing timelike irrotational Killing 
vector field $K^{\alpha}$. The general static spherically symmetric spacetimes in standard schwarzschild coordinates are given by
\begin{eqnarray}
ds^2 =- {\mathcal{B}}(r)\, c^2 \, dt^2 + {\mathcal{A}}(r) \, dr^2 + r^2 d\Omega^2 \, , 
\label{1}
\end{eqnarray}
where, ${\mathcal{A}}(r)$  and ${\mathcal{B}}(r)$ are the generic metric functions and $d\Omega^2 = d\theta^2 +  \sin^2 \theta \, d\phi^2$. 
With ${\mathcal{B}}(r) = 1/{\mathcal{A}}(r) = 1- {2GM}/{c^2 r}$, where $M$ is the gravitational mass of the source, and $c$ the speed of light, Equation (1) furnishes Schwarzschild solution. A wide class of static spherically symmetric spacetime geometries can be represented through Equation (1), notably, Reissner–Nordstr\"om solution, 
Schwarzschild-de Sitter/anti-de Sitter solution, Kiselev BH, Kehagias-Sfetsos solution, Ay\'on-Beato Garc\'ia solution, Weyl-conformal gravity, Janis-Newman-Winicor (JNW) solution describing naked singularity, etc. Many PNPs/modified Newtonian like potentials have been prescribed for many of these static spacetime geometries 
[17,18,19,13,14,9,10,11] however, those modified potentials are unable to reproduce the relativistic features of the corresponding geometries in its entirety with precise accuracy. As the effect of spacetime metrics can only be realized through the dynamics of the particle motion around the 
spacetime geometry, the proposed relativistic analogous potential should contain the information of the velocity of the particle. As we intend to construct our relativistic Newtonian analogous theory from first principle approach, our relativistic Newtonian analogous construct will then be described by an action ${\mathcal{S}}$ and a corresponding Lagrangian ${\mathcal{L}}$, comprising of the information of the test particle motion. We invent a three-dimensional relativistic gravitational action per unit mass in analogous Newtonian framework, carrying the information of the metric functions,  defined by 
\begin{eqnarray}
{\mathcal{S}} =    \int \left[-\frac{e}{2}  - \frac{c^4}{2 e} \, {\mathcal{B}}(r)  
+ \frac{e}{2 c^2} \left(\frac{{\mathcal{A}}(r) \, {\dot r}^2}{{\mathcal{B}}(r)}  
+ \frac{r^2 \,{\dot \Omega}^2}{{\mathcal{B}}(r)}  \right) \right] dt  \, . 
\label{2}
\end{eqnarray}
The corresponding Lagrangian per unit mass is 
\begin{eqnarray} 
{\mathcal{L}} = -\frac{e}{2} - \frac{c^4}{2 e} \, {\mathcal{B}}(r)
+ \frac{e}{2 c^2}  \left(\frac{{\mathcal{A}}(r) \, {\dot r}^2}{{\mathcal{B}}(r)}
+ \frac{r^2 \,{\dot \Omega}^2}{{\mathcal{B}}(r)}  \right) \, . 
\label{3}
\end{eqnarray} 
Here $e$ is an arbitrary constant having the dimension of energy which we will determine from equations of motion, and $c$ is the usual speed of light. $\dot \Omega = \frac{d\Omega}{dt}$, is the angular velocity (in this paper, overdots, denote the derivative with respect to coordinate time ``$t$"). We next obtain the geodesic equations of motion from the Euler-Lagrange equations in spherical geometry, which describe the complete behavior of the test particle dynamics in the presence of gravity defined by our action, given by
\begin{align}
\ddot r &= - \frac{c^6}{2e^2} \frac{{\mathcal{B}}(r)}{{\mathcal{A}}(r)} \frac{\partial {\mathcal{B}}(r)}{\partial r} + \left(\frac{1}{{\mathcal{B}}(r)} \frac{\partial {\mathcal{B}}(r)}{\partial r} - \frac{1}{{\mathcal{A}}(r)} \frac{\partial {\mathcal{A}}(r)}{\partial r} \right) \, \frac{{\dot r}^2}{2} \, 
 + \, \frac{r}{{\mathcal{A}}(r)}\left(1  - 
\, \frac{r}{2{\mathcal{B}}(r)}  \frac{\partial {\mathcal{B}}(r)}{\partial r} \right) \,{\dot \Omega }^2 \, , 
\label{4}
\end{align}
\begin{eqnarray}
\ddot \phi =  -\frac{2 \, \dot r \dot \phi}{r} \left(1-  \frac{r}{{2\mathcal{B}}(r)} \frac{\partial {\mathcal{B}}(r)}{\partial r} \right) \, - \, 2 \, \cot \theta \, \dot \phi \, \dot \theta 
\label{5}
\end{eqnarray}
and 
\begin{eqnarray}
\ddot \theta =  -\frac{2 \, \dot r \dot \theta}{r} \left(1- \frac{r}{2{\mathcal{B}}(r)} \frac{\partial {\mathcal{B}}(r)}{\partial r} \right) \, \, + \, \sin \theta \, \cos \theta \, 
{\dot \phi}^2 \, .
\label{6}
\end{eqnarray}
Integrating the radial geodesic equation of motion given in Equation (4) and using the relation of specific angular momentum $\lambda = \frac{e}{c^2} \frac{r^2 \, \dot \Omega}{{\mathcal{B}}(r)}$,  we obtain the equation for ${\dot r}^2$ (adopting the affine parameter convention used by Misner et al. 1973 [20]), which uniquely describes the test particle dynamics, given by 
\begin{eqnarray}
\frac{e^2}{c^2} \frac{{\mathcal{A}}(r)}{{\mathcal{B}}(r)} \, {\dot r}^2\, + \, c^2 \, {\mathcal{B}}(r) \frac{\lambda^2}{r^2} \, + \, {\mathcal{B}}(r) \, c^4 \ = \, {\mathbb{E}} \, , 
\label{7}
\end{eqnarray}
where ${\mathbb{E}}$ is the integration constant directly associated with the integrals of motion, in the presence of the gravity. In the low energy and test particle motion with small velocity in the Newtonian limit, the integral constant ${\mathbb{E}}$ can be easily identified as the square of the conserved specific energy of the test particle motion (${\mathcal{E}}$), i.e., ${\mathbb{E}} = {\mathcal{E}}^2$, resembling the similar scenario in general relativity in the weak field limit. 
Equivalently, the equation for ${\dot r}^2$ can be obtained by computing the conserved specific Hamiltonian ${\mathcal{H}}$ of the test particle motion using 
Equation (3), given by
\begin{eqnarray}
\frac{e^2}{c^2} \frac{{\mathcal{A}}(r)}{{\mathcal{B}}(r)} \, {\dot r}^2\, + \, c^2 \, {\mathcal{B}}(r) \frac{\lambda^2}{r^2} \, + \, {\mathcal{B}}(r) \, c^4 \, =  e \, (2 {\mathcal{H}} - e ) \, . 
\label{8}
\end{eqnarray}
Comparing Equation (7) and (8), one can find that the constant $e$ is simply equivalent to the conserved specific energy ${\mathcal{E}}$ of the test particle motion. With this, the equation for ${\dot r}^2$ in (7) or in (8) and the 
geodesic equations in (4)-(6) exactly resemble the corresponding equations for the spacetime metrics described by Equation (1). In other words, the action in Equation (2) exactly reproduces the corresponding geodesic equations of motion for the spacetime geometries described by Equation (1). The potential correspond to this Lagrangian could then be easily computed using the most general expression of a relativistic Lagrangian per unit mass in the Newtonian framework which is given by 
\begin{eqnarray}
{\mathcal{L}} = -c^2 \sqrt{1-v^2/c^2} - V \, , 
\label{9}
\end{eqnarray}
where $v = \sqrt{{\dot r}^2 + r^2 {\dot \Omega}^2 }$, the net velocity of the 
test particle, and $V$ is the potential. Equating Equations. (9) and (3), we evaluate the potential function $V$ that can be 
treated as the exact relativistic Newtonian analogous potential corresponding to general static spherically symmetric spacetime geometries, as given by
\begin{align}
V \equiv V_{\rm GN} & = \frac{c^4}{2 e} \, {\mathcal{B}}(r) - \frac{e}{2 c^2} 
\left[  \frac{{\mathcal{A}}(r)}{{\mathcal{B}}(r)}\, \left({\dot r}^2 
+ \frac{1}{{\mathcal{A}}(r)} \, r^2 \, {\dot \Omega}^2 \right)   \right. \nonumber \\
    &\mathrel{\phantom{=}} \left.\kern-\nulldelimiterspace + \frac{2c^4}{e} \sqrt{1-\frac{{\dot r}^2 + r^2 \, {\dot \Omega}^2}{c^2}} -c^2 \right] 
\label{10} \, , 
\end{align} 
where the subscript ``GN" symbolizes ``Geometric-Newtonian". Thus $V_{\rm GN}$ in Equation (10) 
which is a three dimensional relativistic scalar gravitational potential, 
is the relativistic generalization of Newtonian gravitational 
potential, that is consistent with the features of 
corresponding spacetime geometries described by Equation (1). In a real astrophysical 
scenario, the constant ``$e$" in $V_{\rm GN}$ which is equivalent to the specific energy of the particle motion, will be evaluated from the asymptotic boundary condition. 
It is being interestingly found that in the low energy limit of the test particle motion $({e}/c^2 \sim 1)$ or equivalently in the limit ${v^2}/{c^2} <<1$, $V_{\rm GN}$ reduces to the Newtonian like analogous potentials of the corresponding static GR geometries prescribed in [13,14,9,10]. Corresponding to Schwarzschild metric, the exact relativistic Newtonian analogous potential is then given by  
\begin{align}
V_{\rm GN} \bigr\rvert_{\rm SW} &= \frac{c^4}{2 e} \, \left(1-\frac{2GM}{c^2 \, r} \right)  - \frac{e}{2 c^2} \left[\frac{r^2 \,{\dot r}^2}{\left(r-2r_s \right)^2} 
+ \frac{r^3 \,{\dot \Omega}^2 }{r-2r_s}   \right. \nonumber \\
    &\mathrel{\phantom{=}} \left.\kern-\nulldelimiterspace + \frac{2c^4}{e} \sqrt{1-\frac{{\dot r}^2 + r^2 \, {\dot \Omega}^2}{c^2}} -c^2 \right] \, ,
\label{11}
\end{align}
where, $r_s=GM/{c^2}$ and $M$ is the gravitational mass of the source. In the usual low energy limit and in the far field 
approximation ($r >> r_s$), $V_{\rm GN} \bigr\rvert_{\rm SW}$ reduces to that in 
Newtonian gravity. 

\section{Discussion}

For over a century general relativity has successfully withstood all the experimental tests conducted so far, the predictions of general relativity being confirmed in almost all observations and experiments up to the present [2,3]. The recent announcement of the first direct observation and evidence of gravitational waves corresponding to the inspiral and merger of two BHs [21] validates the general relativity also in the strong field region. From the start, there were efforts to check whether the observations concerning gravity can be described by Newtonian like potential. In this regard various corrections have been offered to the Newtonian potential to reproduce the relativistic features of spacetime geometries (see the Introduction) but none of them can describe all the salient relativistic features simultaneously or make it consistent with all the classical experimental tests of gravity. The present work perhaps provides for the first time a method to construct a relativistic Newtonian analogous theory described through a relativistic scalar potential (and the corresponding force law) from a first principle approach for any static spherically symmetric spacetime geometry, starting directly from a generic relativistic gravitational action in analogous to Newtonian framework, 
which gives exactly identical geodesic equations of motion to those of general relativity or of any viable alternative gravitational theory. In constructing this Newtonian analogous theory, no premises (axioms and principles) of geometric theory of gravitation has been used, only the metric function of the spacetime geometry has been taken into the consideration. 

The derived analogous velocity dependent relativistic potential exactly reproduces the relativistic features of the corresponding static spherically symmetric spacetime geometry in its entirety, including all the classical experimental tests of gravity. Several other velocity dependent modified Newtonian potentials corresponding to static spherically symmetric geometries exist in the literature [13,14,9,10]; however, all those potentials have been derived in the low energy limit or equivalently in the limit of ${v^2}/{c^2} << 1$ of the test particle motion. These velocity dependent potentials reproduce with precise accuracy several relativistic features of the corresponding spacetime geometries, mostly in the context of circular orbits, such as location of the photon circular orbit, location of marginally bound and marginally stable circular orbits, radial dependence of angular momentum and mechanical energy of circular orbits but the temporal effects like angular and epicyclic frequencies corresponding to circular orbits can only be reproduced with marginal accuracy; with an error margin of $\sim 6 \%$ in the inner most accretion region, for Schwarzschild case. Time evolution of particles in free fall, also, could not be exactly reproduced by those potentials, although, can provide a good approximation. Moreover, in the context of general orbits, these velocity dependent potentials cannot reproduce the corresponding relativistic orbital dynamics, like angular momentum and energy of the particle, orbital trajectory, pericenter advance, etc., with good precision, unless restricted to the low energy limit. To be more illustrative, in Table 1, we compare the accuracy (maximum percentage error) with which few salient relativistic features are reproduced by existing modified Newtonian potentials or PNPs, and the Newtonian analogous potential prescribed in this work. For simplicity, we choose Schwarzschild geometry.

\begin{table}[ht]
\small
\centerline{\large Table 1}
\centerline{Accuracies of modified Newtonian potentials} 
\centerline{Schwarzschild case} 
\begin{center}
\begin{tabular}{ccccccccccccc}
\hline
\hline
\noalign{\vskip 2mm} 
$\rm Parameters$ &  $\rm V_{PW}$ & $\rm V_{NW}$ &  $\rm V_{TR}$  & $\rm V_{GN}$\\
\hline
\noalign{\vskip 2mm} 
$\rm r_{ph}$  &  $\sim 33.3 \%$ &   $\rm NA$  &  $\rm exact$ & $\rm exact$ \\
\hline
\noalign{\vskip 2mm}
$\rm r_{ms}$  &  $\rm exact$ &  $\rm exact$  &  $\rm exact$  & $\rm exact$\\
\hline
\noalign{\vskip 2mm}
$\rm r_{mb}$  &  $\rm exact$ &   $\rm 13.4 \%$ & $\rm exact$ & $\rm exact$ \\
\hline
\noalign{\vskip 2mm}
$\lambda^{c}$ &   $\sim 6.1 \%$ & $\sim 29.3 \%$  & $\rm exact$  & $\rm exact$\\
\hline
\noalign{\vskip 2mm}
$E^{c}$  & $\sim 12.5 \%$ & $\sim 3.3 \%$ & $\sim 1.75 \%$  & $\rm exact$ \\
\hline
\noalign{\vskip 2mm}
$\Omega^{c}$  &  $\sim 50 \%$  &  $\sim 13.4 \%$  &  $\sim 5.7 \%$ & $\rm exact$ \\
\hline
\noalign{\vskip 2mm}
$\kappa^{c}$  &  $\sim 83.7\%$  &  $\sim 41.4 \%$  &  $\sim 5.7 \%$ & $\rm exact$ \\
\hline
\noalign{\vskip 2mm}
$B.E$  &  $\sim 9.3 \%$  &  $\sim 2.9 \%$  &  $\sim 2.9 \%$  & $\rm exact$\\
\hline
\noalign{\vskip 2mm}
$\mathcal{F}$  &  $\sim 58.9 \%$  &  $\sim 11.1 \%$  &  $\sim 7.3 \%$  & $\rm exact$\\
\hline
\hline
\end{tabular}
\end{center}
\end{table} 

$\rm V_{PW}$, $\rm V_{NW}$, $\rm V_{TR}$ and $\rm V_{GN}$ in Table 1, represent Paczy\'nski-Wiita potential [4], Nowak-Wagnor potential [22], Tejeda-Rosswog potential [13] and Newtonian analogous potential in Equation (11), respectively. $\rm r_{ph}$, $\rm r_{ms}$, $\rm r_{mb}$, represent photon circular orbit, last stable circular orbit, marginally bound orbit, respectively, whereas, $\lambda^{c}$, $E^{c}$, $\Omega^{c}$, $\kappa^{c}$, $B.E$, represent specific angular momentum, specific mechanical energy, angular frequency, radial epicyclic frequency, and binding energy, respectively, at $r \sim r_{ms}$. $\mathcal{F}$ represents radiative flux generated from a standard geometrically thin and optically thick Keplerian accretion disk. 

Being an relativistic Newtonian analogous construct, the relativistic scalar potential prescribed in this work is expected to be useful in analyzing a wide range of complex astrophysical phenomena in strong field gravity. The analytical solution of n-body problem in metric theories remains very difficult, if not elusive. The Newtonian analogous potential should be quite useful in this regard. Most importantly, it is to represent gravitational interaction through a scalar potential by avoiding complex tensorial construct of relativistic gravitation, where by, relativistic astrophysical phenomena, especially in the strong field gravity can be studied accurately in a simple Newtonian framework using the exact Newtonian analogous potential, circumventing complex nonlinear tensorial equations of geometric gravitation. A very appropriate feasible scenario in this regard to use this relativistic potential is to accurately study complex relativistic accretion phenomena around massive compact objects like BH/neutron star/naked singularity or around any other possible spacetime geometry in the Newtonian framework, in strong field gravity. Other situations where Newtonian analogous potential may be useful include tidal disruption of a star by a supermassive BH (SMBH), binary systems or in binary mergers or analyzing galactic dynamics around SMBH, and can even be tested in other applications of gravity as well. This generic Newtonian analogous potential should also be useful to observationally discriminate alternative theories of gravity, that are generally developed in such a manner that they reduce to general relativity in the weak field limit so that the theories automatically satisfy the classical solar system tests of gravitation (though recent observation of gravity waves leaves very little window for the alternative gravitational theories [23]), among themselves and from general theory of relativity in the strong field gravity, and consequently, to test theory of relativistic gravitation in the strong field regime.  



\end{document}